\documentclass[twocolumn,twoside,slac_two]{revtex4}
\usepackage{graphicx}
\usepackage{fancyhdr}
\usepackage{color}
\usepackage{rotating}
\usepackage{subfigure}
\usepackage{enumitem}
\begin{document}

\title{Spectral Studies of Flaring FSRQs at GeV Energies Using Pass 8 \textit{Fermi}-LAT Data}

\author{R. J. Britto, S. Razzaque}
\affiliation{Department of Physics, University of Johannesburg, Auckland Park 2006, South Africa}
\author{B. Lott}
\affiliation{Centre d'\'Etudes Nucl\'eaires de Bordeaux-Gradignan, Universit\'e Bordeaux 1, UMR 5797/IN2P3, 33175 Gradignan, France}
\author{on behalf of the \textit{Fermi}-LAT Collaboration}
\affiliation{}

\begin{abstract}
Flat spectrum radio quasars (FSRQs) are bright active galactic nuclei surrounded by gas clouds within a UV-visible intense radiation field that form the so-called broad line region (BLR). These objects emit relativistic jets from a region close to the central supermassive black hole and through the BLR. The \textit{Fermi}-Large Area Telescope (\textit{Fermi}-LAT) is sensitive to gamma-ray photons from $\sim$30 MeV to more than 300 GeV. We have performed spectral analysis of bright FSRQs in a 5.5 year (2008-2014) data sample collected by \textit{Fermi}-LAT, using the new Pass 8 event selection and instrument response function. Also, our study of flaring episodes in a limited time range brings interesting results while compared to the full 5.5 year data samples.
\end{abstract}

\maketitle

\thispagestyle{fancy}

\section{Modelling the BLR radiation field}

FSRQs constitute a class of active galactic nuclei (AGNs) with a dense BLR in which gamma rays with energy $\gtrsim 10$ GeV are absorbed due to electron-positron pair creation, if produced deep inside the BLR. Indeed, BLR is expected to be denser in FSRQs compared to the BL Lac class.

Operating since 2008, the \textit{Fermi} satellite has amassed more than 6 years of data, continuously surveying the whole sky \cite{Atwood}. The sensitivity of \textit{Fermi}-LAT is ideal for the study of the gamma-ray absorption inside FSRQs in the 100 MeV-300 GeV range. From constraints on gamma-ray absorption we may infer limits on the location of the gamma-ray emission region in the FSRQ jets.

We expect $>$ 10 GeV photons of FSRQs to undergo absorption in the BLR, where the target photon with energy $\epsilon$ is a UV photon from the BLR radiation field. As most of these photons are expected to come from the emission lines, we use a model that includes the 6 strongest lines (NV, Ly$\alpha$, O${\rm VI}$ Ly$\beta$, CIII~NIII, NeVIII~OIV, HeII~Ly$\alpha$) between $\sim$10 to 41 eV \citep{Telfer}. 

We model these lines using a Breit-Wigner distribution, given by:
\begin{equation}
\displaystyle
BW(\epsilon) = \frac {n_{i}~\omega_i} {2~\pi[(\epsilon - \epsilon_{i})^2 + (\omega_i/2)^2]},
\end{equation}

where $n_i$ and $\omega_i$ are the number density and width, respectively, for a given line $i$.

Under the commonly used relations $L_{BLR}= 0.1~L_{disc}$ and $R_{BLR}=\sqrt{L_{disc}}$ \cite{Baldwin,Ghisellini}, where $L_{BLR}$ is the luminosity of the BLR, $R_{BLR}$ its radius, and $L_{disc}$ the luminosity of the accretion disc. The photon density $n_i$ of the radiation field for each line $i$ can be written:
\begin{equation}
\displaystyle
n_{i}[cm^{-3}] \simeq 1.66 \times 10^{11}   \left( \frac{L_{i}}{10^{45}erg~s^{-1}} \right)  \left(\frac{10^{17}cm}{\epsilon_{{i},eV} R_{BLR}^2}\right).
\end{equation}

The opacity is derived from \cite{Gould,Brown} and is expressed as a function of $E$ and $z$:
$$
\displaystyle
\frac{d\tau_{\gamma \gamma}}{dx}(E,z)= \frac{r_0^2}{2} \left[ \frac{m^2c^4}{E(1+z)} \right]^2
$$
\begin{equation}
\displaystyle
\times \sum_{i=1}^6 \left( n_{i} \omega_i \int_{\frac{m^2c^4}{E(1+z)}}^{\infty} \frac{\bar{\varphi} \left[ \frac{\epsilon E(1+z)}{m^2c^4} \right] d\epsilon}{[(\epsilon - \epsilon_{i})^2 + (\omega_i/2)^2] \epsilon^2} \right),
\end{equation}

where $r_0$ is the classical electron radius, $m$ the electron mass.  

The $\tau_{\gamma \gamma}$ opacity in the BLR is then calculated as following:
\begin{equation}
\displaystyle
\tau_{\gamma \gamma} (E,z) = a \times R_{BLR} \times \frac{d\tau_{\gamma \gamma}}{dx} (E,z),
\end{equation}

assuming the gamma rays are produced within $a \times R_{BLR}$, where $R_{BLR}$ is the outer radius of BLR, and $a < 1$. Since this absorption happens at some distance from the supermassive black hole, this corrective factor that we called ``$a$'' represents the fraction of the BLR responsible for the absorption.

In Table \ref{tab:lines} are displayed the line properties of the average spectrum of quasars we used in our model, as they were given in \cite{Telfer}. Since the He II Ly$\alpha$ line has quite large uncertainties, we arbitrary fixed its EW and relative flux to be equal to the ones of N V (uncertainty represented by (*)).

Very high energy gamma rays travelling from far distances undergo absorption in the extragalactic background light (``EBL'', mainly composed of infrared-UV radiation). This absorption is to be considered above 10 GeV and has been implemented in our studies, from the model presented in \cite{Finke}.

Evidence of absorption in the BLR for some FSRQs have been reported in \cite{Poutanen,Stern}.

\begin{figure}[t]
\centering
\includegraphics[width=85mm]{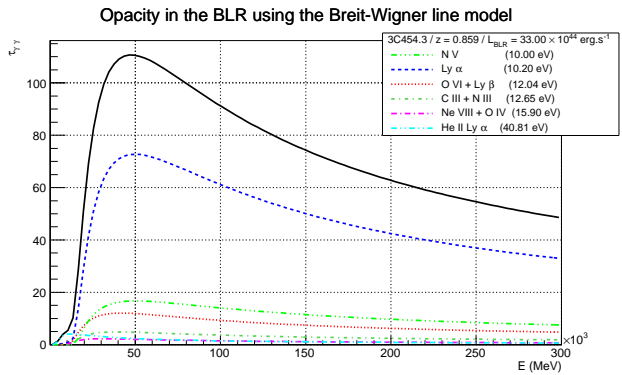}
\caption{Opacity $\tau_{\gamma \gamma}(E,z)$ {\it versus} Energy, for 3C~454.3. The opacity sum on the 6 lines is drawn in plain black.} \label{JACpic2-f1}
\end{figure}
\begin{table}[t]
\begin{center}
\caption{Properties of 5 main lines of the average spectrum of quasars as compiled in \cite{Telfer}, and He II Ly$\alpha$ as we defined it for this study. }
\begin{tabular}{|c|c|c|c|c|c|c|}
\hline
Line             & $\epsilon$ (eV) & EW (eV) & Relative flux\\
\hline
NV              & 10.0 & 0.16 & 0.22\\
Ly$\alpha$      & 10.2 & 0.71 & 1.00\\ 
OVI Ly$\beta$  & 12.04 & 0.19 & 0.191\\ 
CIII NIII      & 12.65 & 0.09 & 0.081\\
NeVIII OIV     & 15.90 & 0.08 & 0.047\\
HeII Ly$\alpha$ & 40.81 & 0.16* & 0.22*\\
\hline
\end{tabular}
\label{tab:lines}
\end{center}
\end{table}

\section{Data processing and model fitting} \label{sec:Data_Proc}

We have analysed data of 7 bright gamma-ray FSRQs. Plots of the spectral energy distributions (SEDs) under the label ``5.5 years'' have been processed from 4 August 2008 till 30 April 2014. The sources we present in this paper are listed in Table \ref{Tab:Sources}, $LII$ and $BII$ being respectively the Galactic latitude and longitude in decimal degrees.

Data were processed using the Pass 8 data representation (P8\_SOURCE\_V4), and the Science Tools version v9-34-01. Signal is reconstructed from each source using the unbinned likelihood tool\footnote{http://fermi.gsfc.nasa.gov/ssc/data/analysis/}, applied to LAT data in the 0.1-300 GeV energy range, within a region of interest (ROI) of 10$^\circ$ radius. A source region extended to an additional 10$^\circ$ annulus accounted for all the point sources of the {\it Second Fermi-LAT source catalog} \cite{2FGL}, and for the Galactic diffuse emission (template\_4years\_P8\_V2\_scaled) and the isotropic diffuse emission (isotropic\_source\_4years\_P8V3).

\begin{figure}[t]
\centering
\includegraphics[width=70mm]{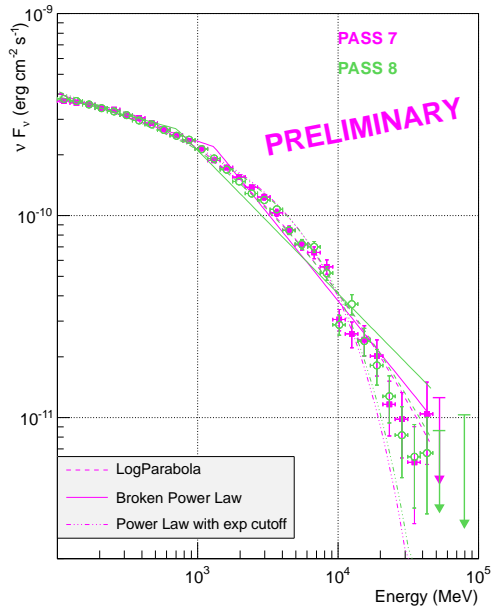}
\caption{Comparison of the SED of 3C~454.3, produced with the Pass 7 (magenta) and Pass 8 (green) data sets.} \label{Fig:3C454_P7_P8}
\end{figure}

We computed the SEDs for all the sources of the selected sample with the Pass 8 data representation. Additionally, for the two brightest objects of our FSRQ sample, {\it i.e.} 3C~454.3 and PKS~1510-089, we also computed the SEDs with the PASS 7 reprocessed dataset (P7REP\_SOURCE\_V15) and verified the consistency of the results with respect to the PASS 8 ones. Although some bin-to-bin fluctuations appear due to energy wise event migrations, the two SEDs (Pass 7 and Pass 8) for both the sources are compatible (as shown in Figure \ref{Fig:3C454_P7_P8} for 3C~454.3).

\begin{table*}[t]
\begin{center}
\caption{Characteristics of the 7 FSRQs used in this paper. Sources are ordered by decreasing flux in the 1-100 GeV energy range. Values of the luminosity in the broad line region ($L_{BLR}$) are taken or derived from \cite{Pacciani, Xiong}.}
\small
\begin{tabular}{c|c|c|c|c|c|c|c|c|c} 
\hline 
Name & RA        & Dec       & LII & BII &  Redshift  & Flux 2FGL     & Flux 1FHL     & Photon index & $L_{BLR}$\\
     & (J2000.0) & (J2000.0) &  (deg.) &  (deg.)    &     $z$      &  1-100 GeV & 10-100 GeV & 1FGL & ($10^{44}erg.s^{-1}$)\\
\hline
3C 454.3         & 22 53 57.7 & +16 08 53.1 & 86.11  & -38.19 &  0.859 & 9.65e-8 & 1.35e-9  & 2.46619 & 33.00\\
PKS 1510-08      & 15 12 50.5 & -09 06 00.9 & 351.28 & 40.13  &  0.360 & 4.06e-8 & 7.35e-10 & 2.40756 & 5.62\\
4C +21.35        & 12 24 54.5 & +21 22 46.9 & 255.08 & 81.65  &  0.434 & 3.54e-8 & 7.43e-10 & 2.54717 & 15.80\\
3C 279           & 12 56 11.0 & -05 47 20.1 & 305.1  & 57.06  &  0.536 & 2.56e-8 & 5.37e-10 & 2.32061 & 3.10\\
PKS 0454-234     & 04 57 03.1 & -23 24 52.0 & 223.7  & -34.9  &  1.003 & 2.27e-8 & 2.99e-10 & 2.20649  & 3.70\\
B2 1520+31       & 15 22 09.8 & +31 44 14.3 & 50.16  & 57.02  &  1.484 & 1.76e-8 & 4.27e-10 & 2.42125 & 8.00\\
PKS (B)1424-418    & 14 27 56.2 & -42 06 18.6 & 321.44 & 17.26  &  1.522 & 1.47e-8 & 2.9e-10  & 2.31004 & 8.91\\
\hline
\end{tabular}
\label{Tab:Sources}
\end{center}
\end{table*}

A first set of fits was performed from 100 MeV till the highest energy data point (excluding upper limits), using a log-parabola (LP: $dN(E)/dE=N_0~(E/E_0)^{-\alpha-\beta~log(E/E_0))}$, with $E_0$ kept fixed at 297.6 MeV, and where ``log'' is the natural logarithm), a broken power law (BPL: $dN(E)/dE=N_0~(E/E_b)^{-\Gamma_i}$, with $i=1$ if $E<E_b$ and $i=2$ if $E>E_b$), and a power law with an exponential cutoff (PLEC: $dN(E)/dE=N_0~(E/E_p)^{-\Gamma_{PLEC}}~exp(-E/E_c)$, with $E_p$ kept fixed at 412.7 MeV). Other sets of fits were performed by adding exponential factors to model the EBL absorption \cite{Finke} and the opacity of the BLR.

As the fits presented in the Sections \ref{Sec:5.5years} and \ref{sec:Flares} are binning dependant, the values of the fit parameters vary from one choice of binning to another. Narrow data binning could held spurious fluctuations, while wide data-binning could hide features. In order to estimate this systematic effect, we do the following: a first LP fit is performed on the SED, while keeping all parameters fixed to the values obtained by the unbinned likelihood analysis, and a $\chi^2/ndf$ is returned. A second LP fit is performed with $N_0$, $\alpha$ and $\beta$ kept free, $E_b$ kept fixed to the value returned by the unbinned likelihood analysis. We compare the fits to validate that the binned fit is compatible with the unbinned fit, though results differ. These systematics could be overcome by implementing an unbinned analysis for all fitted models in future.

Our modelling study is then done by fitting the SEDs using the LP, BPL and PLEC function that all now include EBL. These fits will be reported in the Sections \ref{Sec:5.5years} and \ref{sec:Flares}, and compared to the fits that include both EBL and BLR absorption. The latter fits are written as $LP\tau$, $BPL\tau$, and $PLEC\tau$.

The observed spectrum $F_{obs}(E)$ will then be ultimately written:
\begin{equation}
\displaystyle
F_{obs}(E)=e^{-\tau_{EBL}(E,z)}~e^{-a~\tau_{\gamma \gamma}(E,z)}~F_{int}(E),
\end{equation}

where $F_{int}(E)$ is the LP, BPL or PLEC fitting function. Parameter $a$ is kept free in the [$10^{-5}$, 1] range to account for the fraction of radius of the BLR in which gamma rays may be absorbed.

The fitting procedure using the absorption models is implemented by interpolation of both the $\tau_{EBL}$ and $\tau_{\gamma \gamma}$ graphs. While comparing each ``EBL + BLR absorption'' fit (LP$\tau$, BPL$\tau$ or PLEC$\tau$) with the ``EBL + no BLR absorption'' fit (LP, BPL or PLEC), and if both fits have a $\chi^2/ndf \lesssim 1$, we obtain a p-value which indicates the discrepancy between the fit with model and the fit without model, for a given function.


\section{Results on the 5.5 years of data} \label{Sec:5.5years}

\begin{figure*}[t]
\centering
\subfigure{\includegraphics[width=65mm]{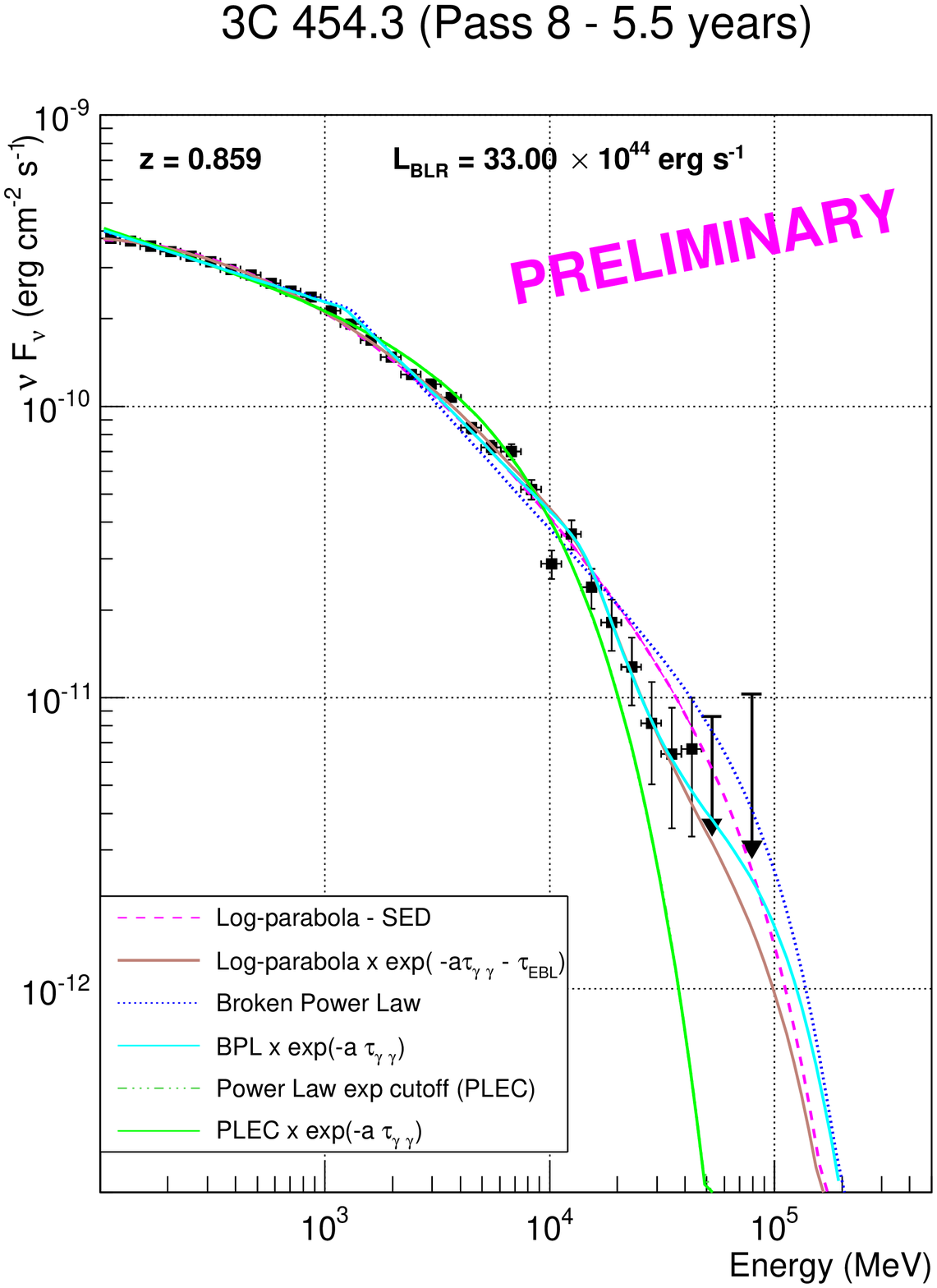}}
\subfigure{\includegraphics[width=65mm]{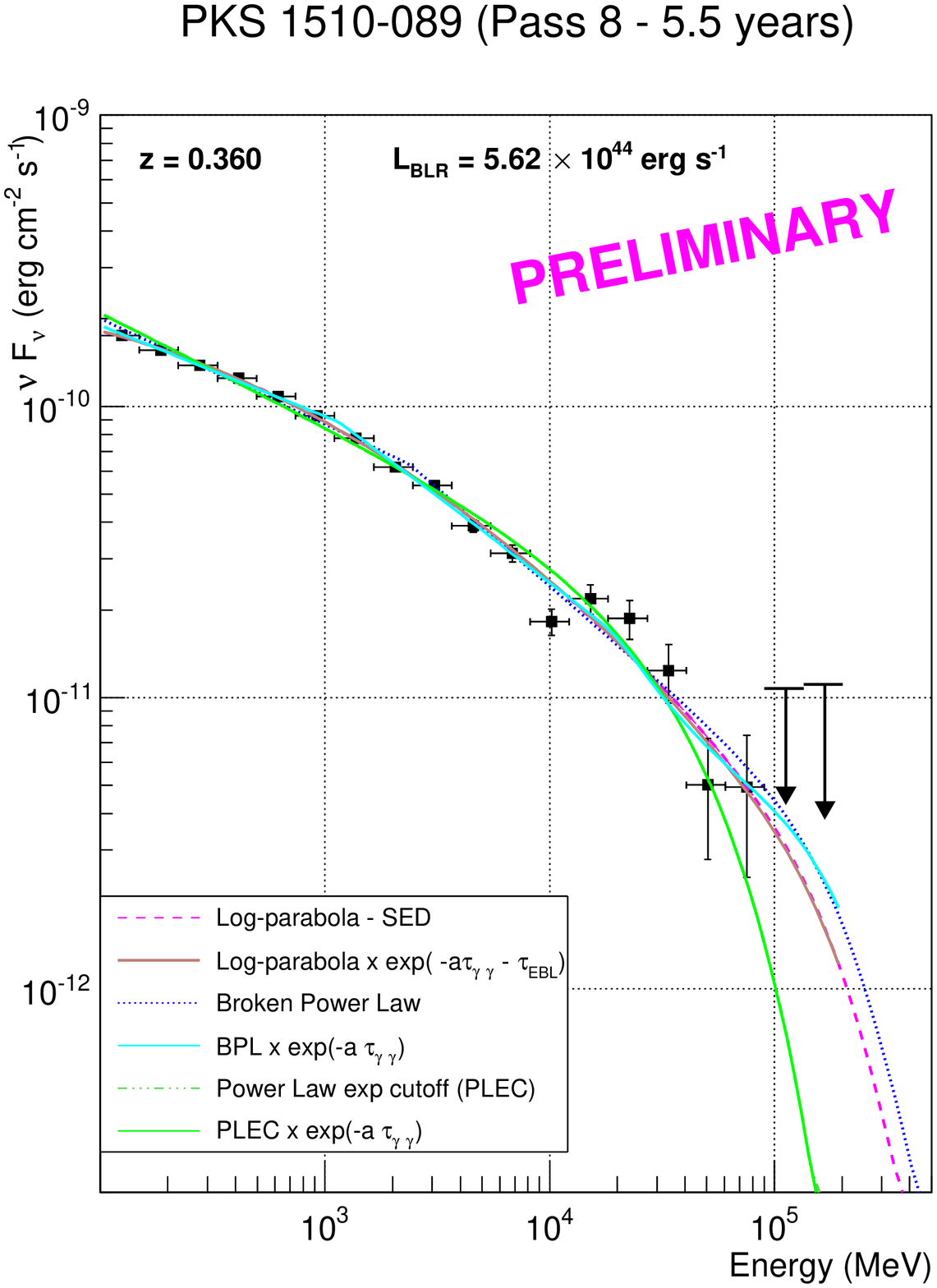}}\\
\subfigure{\includegraphics[width=65mm]{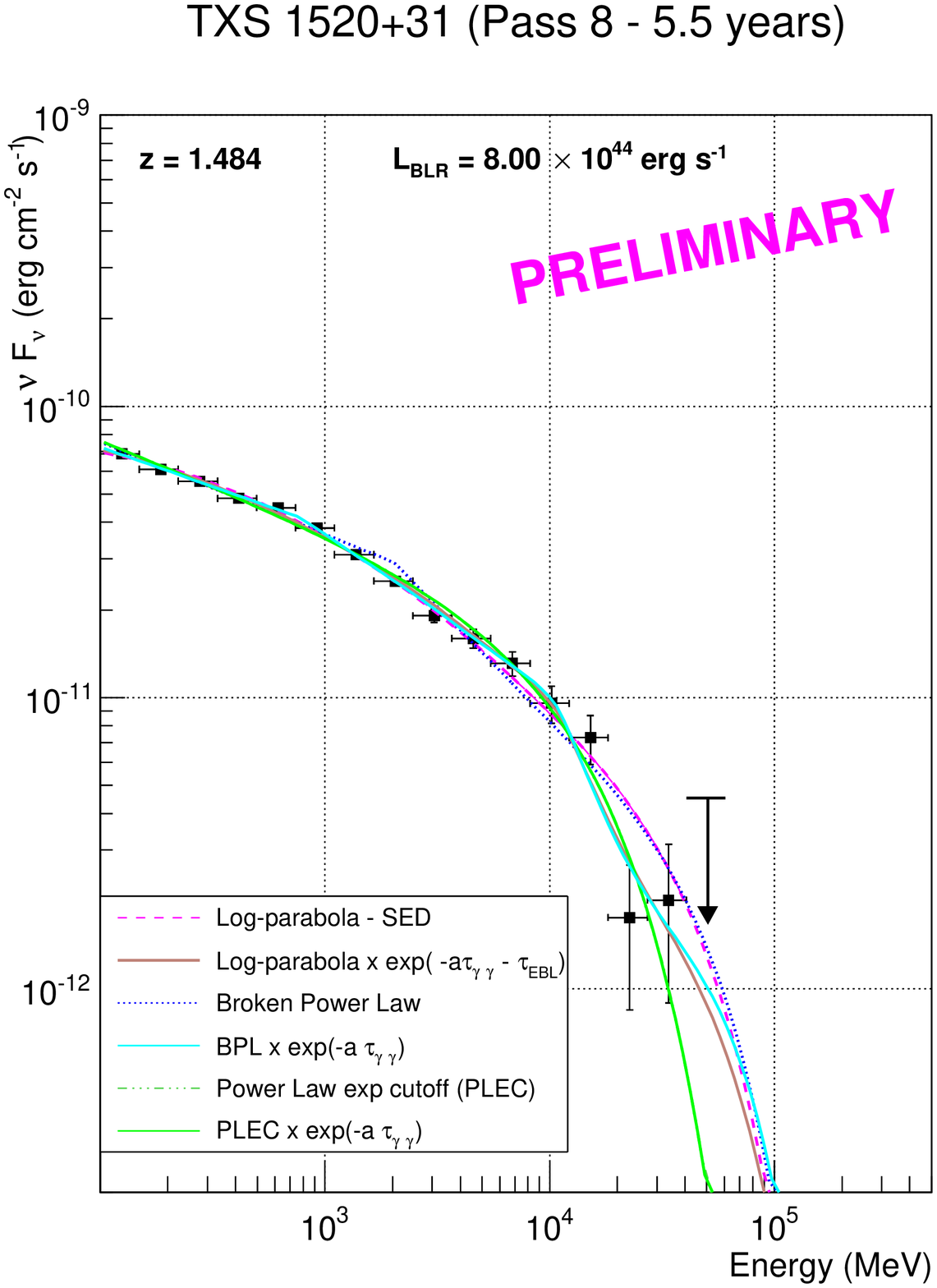}}
\subfigure{\includegraphics[width=65mm]{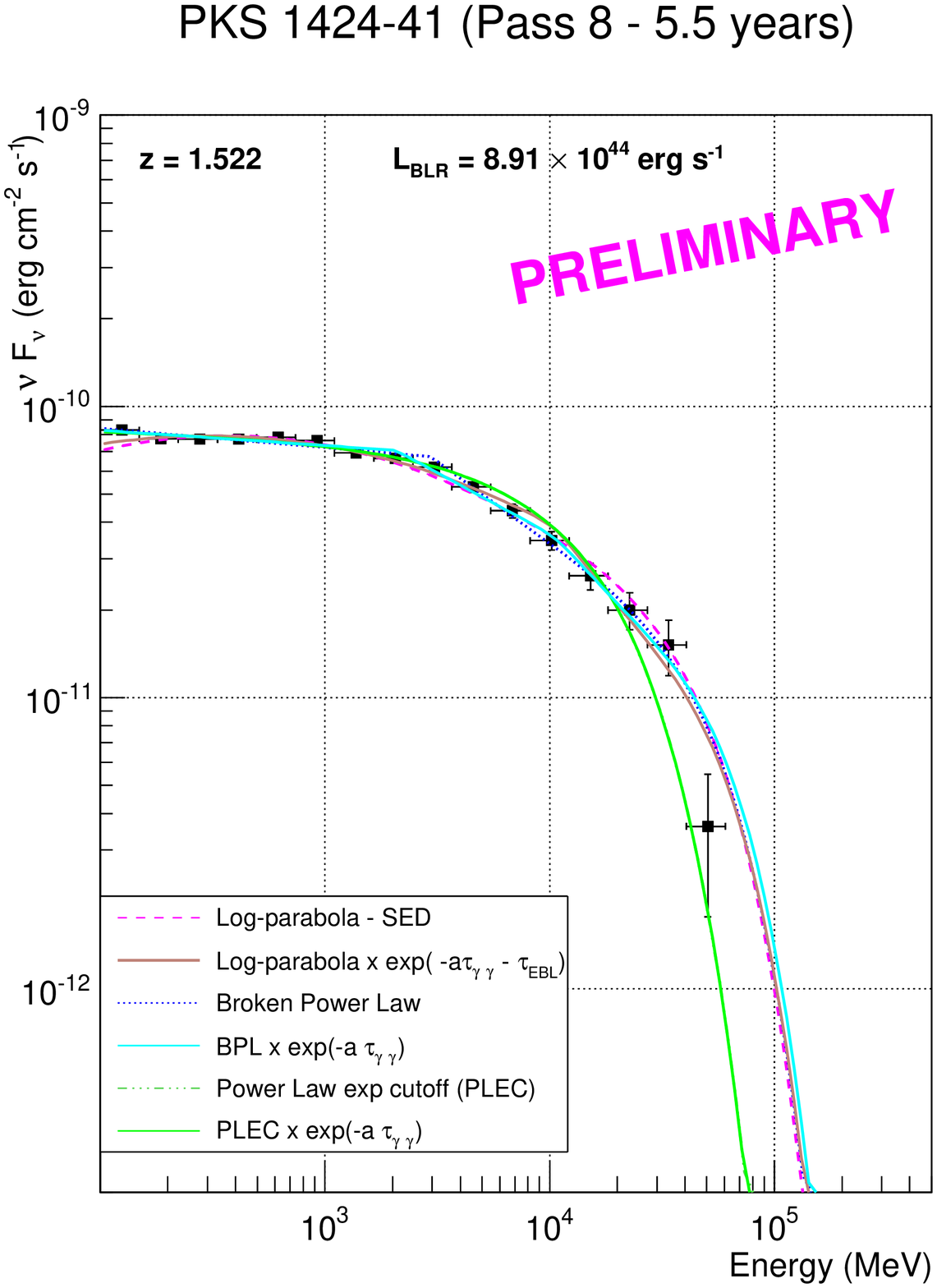}}
\caption{SEDs of 3C~454.3, PKS~1510-089, TXS~1520+31 and PKS~1424-41. $BPL$ and $PLEC$ fits are often hidden beneath $BPL\tau$ and $PLEC\tau$.} \label{Fig:5.5years}
\end{figure*}

We present now the results on the 7 bright FSRQs during the 5.5 years period previously defined in Section \ref{sec:Data_Proc}. Data under the label ``5.5 years'' are processed from 4 August 2008 till 30 April 2014. In Figure \ref{Fig:5.5years} are presented the SEDs of 3C~454.3, PKS~1510-089, TXS~1520+31 and PKS~1424-41. Some of the fits are not visible, mainly the BPL and PLEC. The reason is that they are overwritten by the BPL$\tau$ and PLEC$\tau$ functions, for which the parameter $a$ is very small.

\begin{table*}[t]
\begin{center}
\caption{Fitting parameters and derived significances for the 7 bright FSRQs of our sample. We mention (*) when $a$ reached the lower edge of the fitting interval.}
\scriptsize
\rotatebox{90}{
\begin{tabular}{c|ccccccccc}
\hline
Parameter   &   Model &    & 3C 454.3  & PKS 1510-089 & 4C +21.35 & 3C 279 & PKS 0454-234 & TXS 1520+31 & PKS1424-24\\
 \hline
                    & LP$\_unbinned$ & $\alpha$    & 2.248   $\pm$ 0.005   &   2.294  $\pm$   0.008  &  2.240  $\pm$   0.009  &  2.254  $\pm$  0.010   &  2.067  $\pm$  0.013   &  2.273   $\pm$  0.013  &   2.090   $\pm$ 0.008   \\
                    & LP$\_unbinned$ & $\beta$     & 0.0897  $\pm$ 0.0024  &  0.0513  $\pm$  0.0036  & 0.0445  $\pm$  0.0044  & 0.0588  $\pm$ 0.0053   & 0.0793  $\pm$ 0.0053   & 0.0689   $\pm$ 0.0059  &  0.0753   $\pm$ 0.0039  \\
                    & PL	     & $E_{b}$     & 298     $\pm$  4.4    &	 258  $\pm$   6.4    &    317  $\pm$  10.8    &    284  $\pm$ 10.2     &    358  $\pm$ 16.9     &    281   $\pm$ 11.2	 &     620   $\pm$ 23.8    \\
function            & LP	     & $\alpha$    & 2.240   $\pm$ 0.008   &   2.292  $\pm$   0.030  &  2.274  $\pm$   0.025  &  2.266  $\pm$  0.027   &  2.070  $\pm$  0.019   &  2.274   $\pm$  0.026  &   2.077   $\pm$  0.015  \\
{\it  (without BLR} & LP	     & $\beta$     & 0.0980  $\pm$ 0.0039  &  0.0511  $\pm$  0.0093  & 0.0404  $\pm$  0.0092  & 0.0539  $\pm$ 0.0099   & 0.0761  $\pm$ 0.0087   & 0.0687   $\pm$ 0.0110  &  0.0713   $\pm$ 0.0088  \\
{\it  absorption}   & BPL	     & $\Gamma_1$  & 2.242   $\pm$ 0.345   &   2.363  $\pm$   0.030  &  2.280  $\pm$   0.028  &  2.332  $\pm$  0.029   &  2.138  $\pm$  0.016   &  2.317   $\pm$  0.033  &   2.066   $\pm$  0.010  \\
{\it   model)}      & BPL	     & $\Gamma_2$  & 2.856   $\pm$ 2.108   &   2.673  $\pm$   0.075  &  2.521  $\pm$   0.057  &  2.690  $\pm$  0.098   &  2.681  $\pm$  0.064   &  2.791   $\pm$  0.121  &   2.564   $\pm$  0.087  \\
                    & BPL	     & $E_{b}$    & 1307    $\pm$ 5211.8  &	2387  $\pm$  971.5   &   1439  $\pm$  573.3   &   2738  $\pm$ 979.1    &   3062  $\pm$   0.3    &   2036   $\pm$ 674.1   &    2947   $\pm$ 573.6   \\
                    & PLEC	     & $\Gamma_{PLEC}$    & 2.241   $\pm$ 0.016   &   2.386  $\pm$   0.034  &  2.340  $\pm$   0.035  &  2.324  $\pm$  0.032   &  2.088  $\pm$  0.024   &  2.308   $\pm$  0.038  &   2.028   $\pm$  0.013  \\
                    & PLEC	     & $E_{c}$     & 8250    $\pm$ 721.2   &   40228  $\pm$ 13971.0  &  52905  $\pm$ 28547.3  &  26932  $\pm$ 8335.2   &  14581  $\pm$ 2929.5   &  14378   $\pm$ 3950.2  &   16459   $\pm$ 2097.1  \\
\hline
                   &  LP$\tau$        & $\alpha$    &  2.243  $\pm$ 0.008   &   2.294  $\pm$   0.033  &  2.274  $\pm$   0.025  &  2.268  $\pm$  0.029   &  2.082  $\pm$  0.017   &  2.289   $\pm$  0.033  &   2.079   $\pm$  0.013  \\
                   &  LP$\tau$        & $\beta$	   & 0.0876  $\pm$ 0.0051  &  0.0498  $\pm$  0.0117  & 0.0404  $\pm$  0.0092  & 0.0526  $\pm$ 0.0130   & 0.0590  $\pm$ 0.0103   & 0.0506   $\pm$ 0.0162  &  0.0542   $\pm$ 0.0142  \\
function           &  BPL$\tau$       & $\Gamma_1$  & 2.250   $\pm$ 0.012   &   2.312  $\pm$   0.037  &  2.251  $\pm$   0.031  &  2.301  $\pm$  0.031   &  2.006  $\pm$  0.025   &  2.272   $\pm$  0.045  &   2.052   $\pm$  0.011  \\
{\it  (with BLR}   &  BPL$\tau$       & $\Gamma_2$  & 2.738   $\pm$ 0.049   &   2.583  $\pm$   0.056  &  2.507  $\pm$   0.048  &  2.590  $\pm$  0.109   &  2.283  $\pm$  0.034   &  2.518   $\pm$  0.069  &   2.403   $\pm$  0.066  \\
{\it  absorption}  &  BPL$\tau$       & $E_{b}$	   & 1225    $\pm$ 142.9   &	1152  $\pm$    372.0 &   1062  $\pm$	379.5 &   1616  $\pm$   743.9  &    544  $\pm$   101.1  &    769   $\pm$   315.1 &    2052   $\pm$     1.2 \\
{\it   model)}     &  PLEC$\tau$      & $\Gamma_{PLEC}$	   & 2.241   $\pm$ 0.016   &   2.386  $\pm$   0.034  &  2.341  $\pm$   0.035  &  2.324  $\pm$  0.032   &  2.088  $\pm$  0.024   &  2.308   $\pm$  0.038  &   2.028   $\pm$  0.013  \\
                   &  PLEC$\tau$      & $E_{c}$	   & 8251    $\pm$ 721.4   &   40239  $\pm$  13864.7 &  52946  $\pm$  27502.4 &  26936  $\pm$  8309.2  &  14582  $\pm$  2917.7  &  14381   $\pm$  3982.3 &   16463   $\pm$  2088.6 \\
\hline
                  &   LP$\tau$       &             & \bf \color{blue} 0.00516 $\pm$ 0.00185 &  0.00147 $\pm$  0.01118 &  0.00001 $\pm$ 0.00139 &  0.00232 $\pm$ 0.02143 &  0.02140 $\pm$ 0.01007 &  0.01500 $\pm$ 0.00943 &   0.00745 $\pm$  0.00515 \\
       a          &   BPL$\tau$      &             & 0.00666 $\pm$ 0.00195 &  0.00785 $\pm$  0.00735 &  0.00001 $\pm$ 0.00167 &  0.00919 $\pm$ 0.01496 &  0.03641 $\pm$ 0.00891 &  \bf \color{blue} 0.02002 $\pm$ 0.00881 &   0.00558 $\pm$  0.00488 \\
                  &   PLEC$\tau$     &             & 0.00001 $\pm$ 0.00055 &  0.00001 $\pm$  0.00529 &  0.00001 $\pm$ 0.00151 &  0.00001 $\pm$ 0.00682 &  0.00001 $\pm$ 0.00606 &  0.00001 $\pm$ 0.01584 &   0.00001 $\pm$  0.00248 \\
\hline
                  &  LP$\_unbinned$  &             & 36.419    (30)  	   &   7.454	     (17)    &  18.821         (16)   &  5.288       (16)      & 14.783 	(16)    &  8.988	 (15)    &  36.173	   (16)   \\
 $\chi^2(ndf)$    &    LP            &             & \bf \color{blue} 29.316 (27)  & 7.278   (14)    &  13.009         (13)   &  4.609       (13)      & 11.516 	(13)    &  \bf \color{blue} 6.448 (12) & 22.826	   (13)   \\
{\it without abs.} &  BPL             &             & 45.647    (26)  	   &   9.963	     (13)    &  11.269         (12)   &  7.290       (12)      & 28.960 	(12)    &  8.826	 (11)    &  19.366	   (12)   \\
  {\it model}      &  PLEC            &             & 42.999    (26)  	   &  13.115	     (13)    &  21.817         (12)   &  7.422       (12)      & 17.094 	(12)    &  5.158	 (11)    &  17.037	   (12)   \\
\hline
  $\chi^2(ndf)$   &  LP$\tau$        &             & \bf \color{blue} 17.591 (26)  & 7.245   (13)    &  13.017         (12)   &  4.582       (12)      &  6.526 	(12)    &  3.181	 (11)    &  20.495	   (12)   \\
  {\it with abs.} &  BPL$\tau$       &             & 25.071    (25)  	   &   7.242	     (12)    &   9.870         (11)   &  5.694       (11)      &  4.282 	(11)    &  \bf \color{blue} 2.069 (10)    &  17.474	   (11)   \\
    {\it model}   &  PLEC$\tau$      &             & 43.017    (26)  	   &  13.117	     (13)    &  21.823         (12)   &  7.424       (12)      & 17.095 	(12)    &  5.159	 (11)    &  17.042	   (12)   \\
\hline
 $\Delta \chi^2$ &  LP/PL$\tau$     &             & \bf \color{blue} 1.173e+01	 & 3.360e-02	     &     7.208e-03	      &    2.747e-02	       &    4.989e+00	        &    3.267e+00           &     2.331e+00	  \\
        (ndf)     &  BPL/BPL$\tau$   &             & 2.058e+01       	   &	 2.720e+00	     &     1.399e+00	      &    1.596e+00	       &    2.468e+01	        &   \bf \color{blue}  6.757e+00 &     1.892e+00	  \\
                  &  PLEC/PLEC$\tau$ &             & 1.812e-02       	   &	 1.888e-03	     &     6.614e-03	      &    1.464e-03	       &    1.646e-03	        &    6.277e-04           &     4.025e-03	  \\
\hline
                 &   LP/LP$\tau$    &             & \bf \color{blue} 6.166e-04 & 8.546e-01	     &     9.323e-01	      &    8.684e-01	       &    2.551e-02	        &    7.070e-02           &     -	  \\
    p-value       &  BPL/BPL$\tau$   &             &    -        	   &	 9.908e-02	     &     2.369e-01	      &    2.064e-01	       &    -    	        &  \bf \color{blue} 9.337e-03 &     -	  \\
                  &  PLEC/PLEC$\tau$ &             &    -       	   &	 -      	     &     -     	      &   -     	       &    -    	        &   -                    &     -  	  \\
\hline
\end{tabular}}
\label{Tab:5.5years}
\end{center}
\end{table*}

We would consider having evidence for absorption in the BLR if we get all of the following:

\begin{itemize}
\item at least one good quality fit among one of the fits with EBL+BLR absorption (LP$\tau$, BPL$\tau$ and PLEC$\tau$);
\item parameter ``a'' with a relatively small error bar;
\item a small p-value or a bad fit of the corresponding function with only EBL absorption (LP, BPL or PLEC).
\end{itemize}

In Table \ref{Tab:5.5years} are displayed the fit parameters of the 7 sources, along with the p-values used to compare the models (with {\it versus} without absorption). In blue bold face are the parameters that suggest a possible BLR absorption, as some of the above conditions are partially met for 3C~454.3 (with a p-value of $6.2 \times 10^{-4}$ / 3.9 $\sigma$ C.L.), and for TXS 1520+31 (with a p-value of $9.3 \times 10^{-3}$ / 2.6 $\sigma$ C.L.). We have no hint of absorption for the other sources we studied.


\section{Results on high state/flaring episodes} \label{sec:Flares}

\begin{figure*}[t]
\centering
\subfigure{\includegraphics[width=56mm]{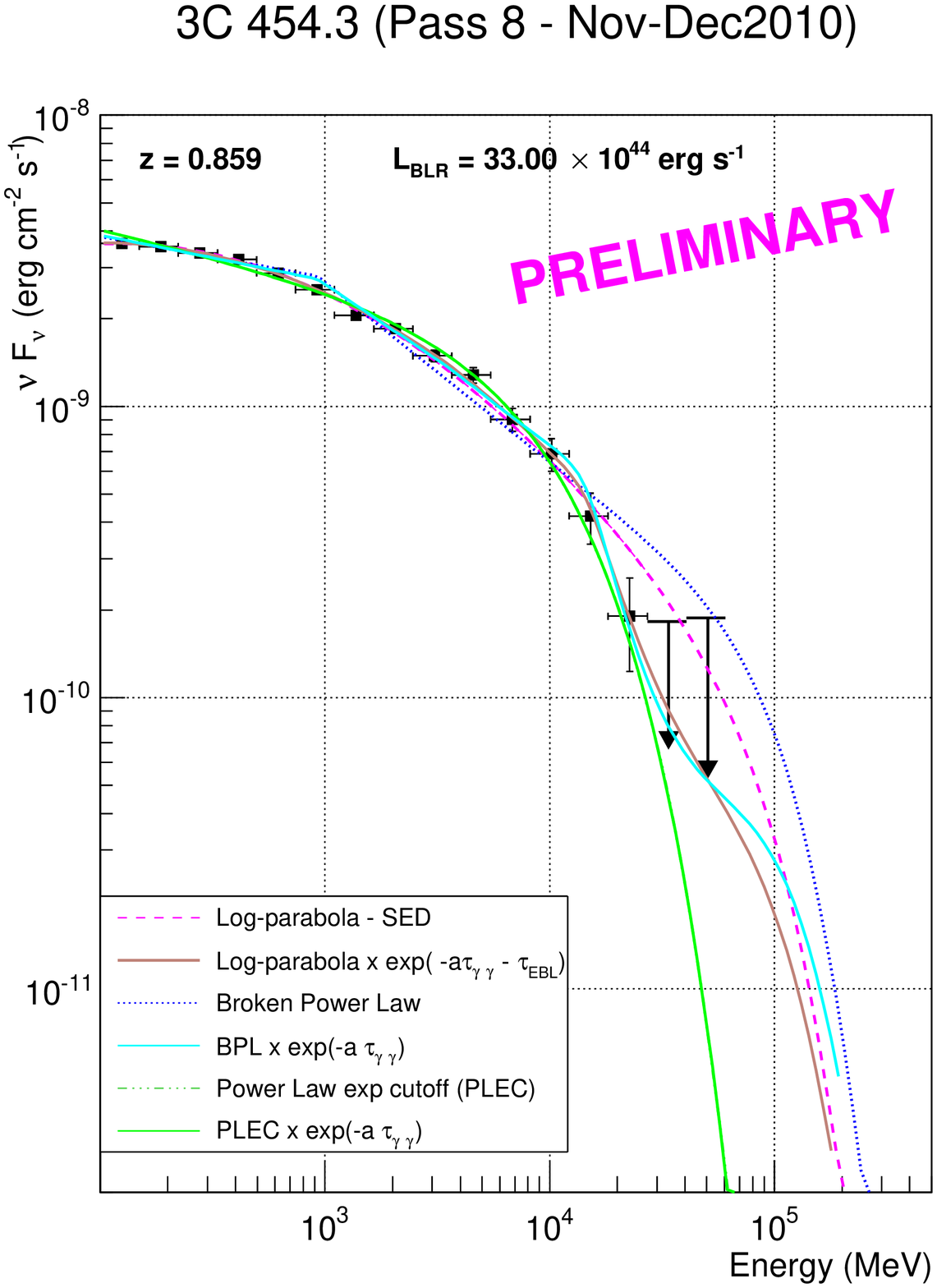}}
\subfigure{\includegraphics[width=56mm]{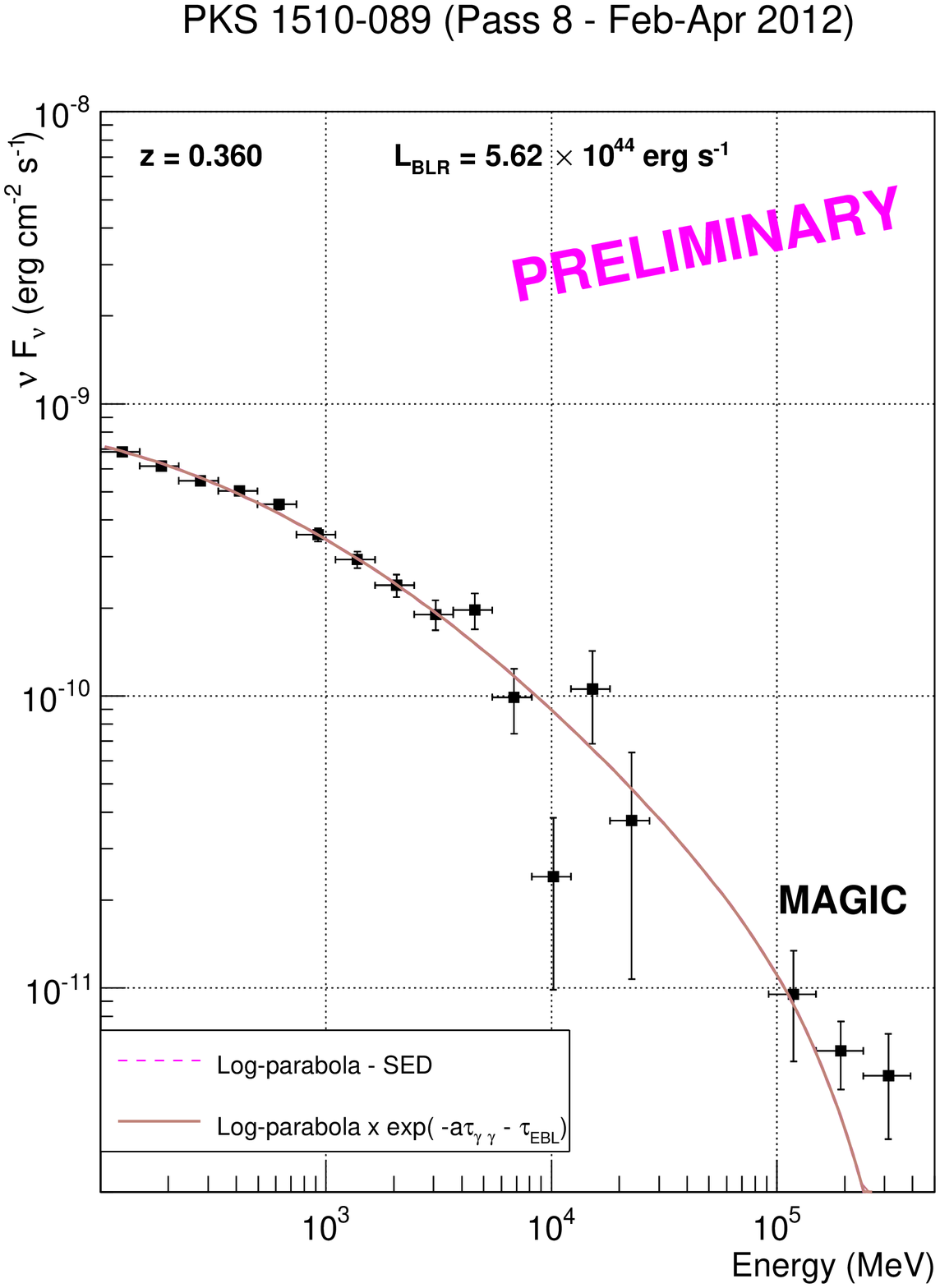}}
\subfigure{\includegraphics[width=56mm]{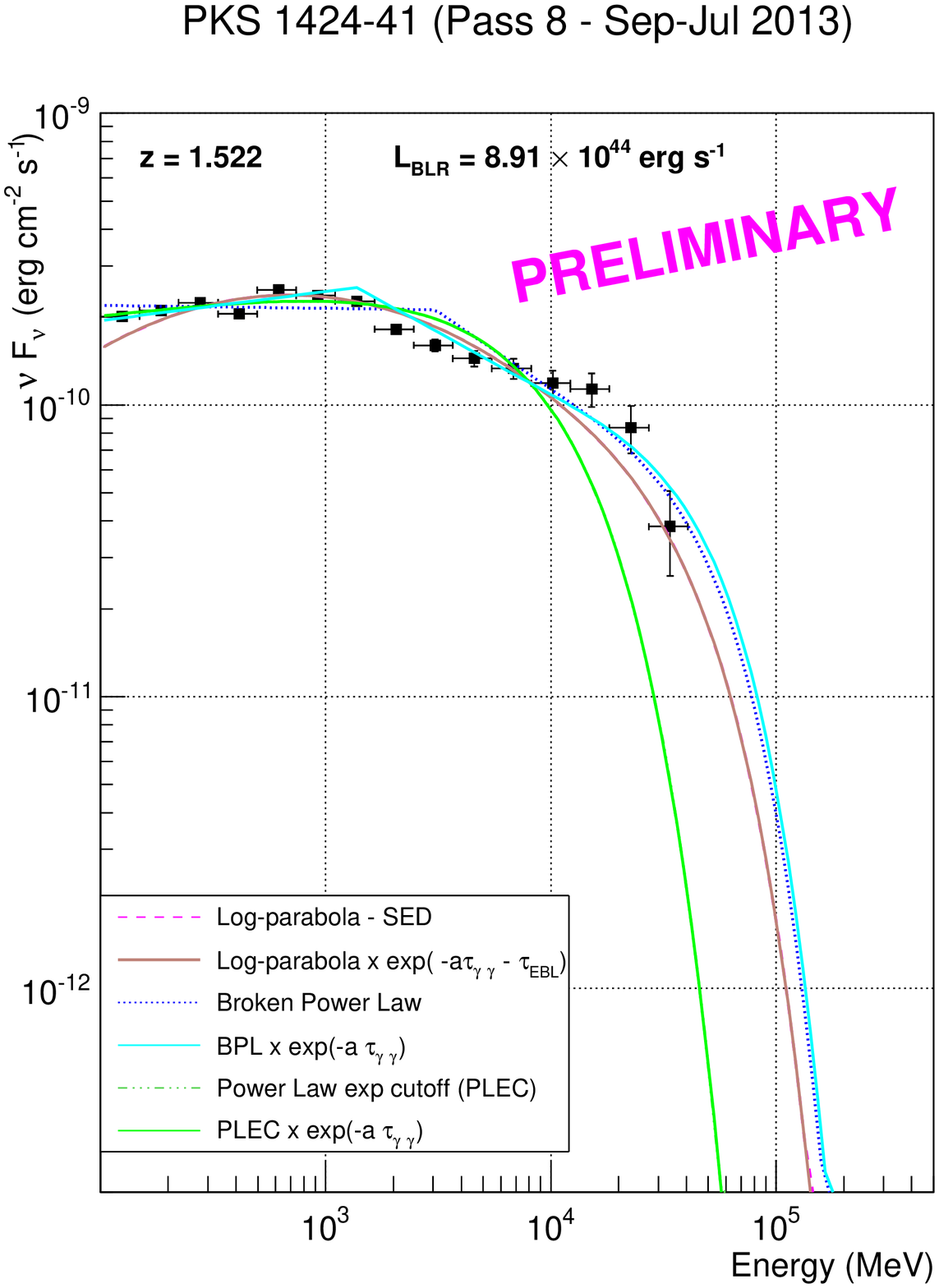}}
\caption{SEDs of 3C~454.3, PKS~1510-089 and PKS~1424-41 during the outburst periods. $PLEC$ fits are hidden beneath $PLEC\tau$ for 3C~454.3 and PKS~1424-41. The flare of PKS~1510-089 was studied along with MAGIC data above $\sim$90 GeV and the combined LAT-MAGIC SED was fitted only with LP and LP$\tau$ function (LP hidden beneath LP$\tau$).} \label{Fig:Flares}
\end{figure*}

Data were analysed during flaring/high state periods for the following sources:

\begin{itemize}
\item 3C~454.3 (high state and giant flare) during 02 Nov-05 Dec 2010 (MJD 55502.5-55535.5)\cite{Abdo2011};
\item PKS~1510-089 during 19 Feb-04 Apr 2012 (MJD 55976.0-56021.0), along with MAGIC data from the same period (MAGIC data points taken from \cite{Aleksic});
\item PKS~1424-41 during 30 Sep 2012-27 Jul 2013 (MJD 56200.0-56500.0), as a combined series of 4 successive radio flares).
\end{itemize}

\begin{table*}[t]
\begin{center}
\caption{Fitting parameters and derived significances for the 3 flares we studied in our FSRQ sample. We mention (*) when $a$ reached the lower edge of the fitting interval.}
\begin{tabular}{c|ccccc}
\hline
Parameter   &   Model &    & 3C 454.3  & PKS 1510-089  & PKS1424-24\\
            &         &    &  (Nov-Dec 2010) & (Feb-Apr 2012) & (Sep 2012-Jul 2013)\\
 \hline
                    & LP$\_unbinned$ & $\alpha$    &  2.152   $\pm$  0.008    & < 2.268 	0.027  &  2.022   $\pm$   0.010     \\
                    & LP$\_unbinned$ & $\beta$     & 0.0895   $\pm$  0.0036   &  0.0451        0.0108  & 0.0766   $\pm$ 0.0052      \\
                    & PL	     & $E_{b}$     & 	286   $\pm$   8.9     &     260 	6.0    &    940   $\pm$ 35.4	    \\
function            & LP	     & $\alpha$    &  2.153   $\pm$   0.016   &   2.299 	0.040  &  2.069   $\pm$  0.014      \\
{\it  (without BLR} & LP	     & $\beta$     & 0.0879   $\pm$  0.0088   &  0.0569        0.0117  & 0.1134   $\pm$ 0.0120      \\
{\it  absorption}   & BPL	     & $\Gamma_1$  &  2.140   $\pm$   0.022   &   2.367 	0.038  &  2.010   $\pm$  0.003      \\
{\it   model)}      & BPL	     & $\Gamma_2$  &  2.617   $\pm$   0.036   &   2.867 	0.076  &  2.530   $\pm$  0.055      \\
                    & BPL	     & $E_{b}$    & 	921   $\pm$    0.4    &    2603        912.7   &   3062   $\pm$   1.6	    \\
                    & PLEC	     & $\Gamma_{PLEC}$    &  2.183   $\pm$   0.034   &   2.269 	0.049  &  1.903   $\pm$  0.021      \\
                    & PLEC	     & $E_{c}$     &   9981   $\pm$  2111.4   &    7921        2586.5  &   8361   $\pm$ 1117.7      \\
\hline
                   &  LP$\tau$        & $\alpha$   &  2.157   $\pm$   0.018   &   2.299 	0.022  &  2.069   $\pm$  0.014      \\
                   &  LP$\tau$        & $\beta$	   & 0.0764   $\pm$  0.0116   &  0.0569        0.0065  & 0.1134   $\pm$ 0.0120      \\
function           &  BPL$\tau$       & $\Gamma_1$ &  2.156   $\pm$   0.029   &       -                &  1.900   $\pm$  0.019      \\
{\it  (with BLR}   &  BPL$\tau$       & $\Gamma_2$ &  2.518   $\pm$   0.050   &       -                  &  2.426   $\pm$  0.036      \\
{\it  absorption}  &  BPL$\tau$       & $E_{b}$	   & 	921   $\pm$	 0.2  &        -                 &   1375   $\pm$     0.2     \\
{\it   model)}     &  PLEC$\tau$      & $\Gamma_{PLEC}$   &  2.183   $\pm$   0.034   &        -                 &  1.903   $\pm$  0.021      \\
                   &  PLEC$\tau$      & $E_{c}$	   &   9983   $\pm$   2113.2  &       -                  &   8361   $\pm$  1135.7     \\
\hline
                  &   LP$\tau$       &             &  0.00703 $\pm$  0.00551  &  0.04667         0.04550 &  0.00001 $\pm$   0.00132  \\
       a          &   BPL$\tau$      &             &  \bf \color{blue} 0.01059 $\pm$  0.00591  &       -                   &  0.00001 $\pm$   0.00320  \\
                  &   PLEC$\tau$     &             &  0.00001 $\pm$  0.00629  &      -                    &  0.00001 $\pm$   0.00108  \\
\hline
                  &  LP$\_unbinned$  &             &  4.553	    (14)      &  67.828 	(17)       & 79.634         (15)	  \\
 $\chi^2(ndf)$    &    LP            &             &  4.041	    (11)      &  20.807 	(14)       & 37.481         (12)	  \\
{\it without abs.} &  BPL             &            & \bf \color{blue} 13.158	    (10)      &  15.137 	(13)       & 101.358         (11)	  \\
  {\it model}      &  PLEC            &            &  6.912	    (10)      &  37.461 	(13)       & 77.643         (11)	  \\
\hline
  $\chi^2(ndf)$   &  LP$\tau$        &             &  0.928	    (10)      &  20.807         (13)       & 37.489	   (11)       \\
  {\it with abs.} &  BPL$\tau$       &             &  \bf \color{blue} 3.780	    ( 9)      &       -                    & 24.648	   (10)       \\
    {\it model}   &  PLEC$\tau$      &             &  6.914	    (10)      &       -                    & 77.652	   (11)       \\
\hline
 $\Delta \chi^2$  &  LP/PL$\tau$     &             & 	3.113e+00	      &   6.395e-11 &    7.583e-03	      \\
        (ndf)     &  BPL/BPL$\tau$   &             & 	\bf \color{blue} 9.378e+00	      &       -                    &    7.671e+01	      \\
                  &  PLEC/PLEC$\tau$ &             & 	1.586e-03	      &       -                    &    9.270e-03	      \\
\hline
                  &   LP/LP$\tau$    &             & 	7.769e-02	      &       -                    &    -	      \\
    p-value       &  BPL/BPL$\tau$   &             & 	\bf \color{blue} 2.195e-03	      &      -                     &    -	      \\
                  &  PLEC/PLEC$\tau$ &             & 	0.000e+00	      &      -                     &    -	      \\
\hline
\end{tabular}
\label{Tab:Flares}
\end{center}
\end{table*}

During these outburst episodes, the gamma-ray emission region can have a different location compared to the quiescent state. In this section we present the results obtained on these three high states (Figure \ref{Fig:Flares} and Table \ref{Tab:Flares}). Under the same criteria than the ones used in Section \ref{Sec:5.5years}, we still find no evidence of BLR absorption, though we still have a hint of it for 3C~454.3 with a p-value of around $2.2 \times 10^{-3}$ for the discrepancy between the BPL and BPL$\tau$ fits (significance of about 3 $\sigma$). Due to the unusual shape of the SED of PKS~1424-41 during this series of 4 flares, all fits have a large $\chi^2$ value.
 
Though we dispose of less photons in the data analysis of flaring episodes, during strong and long flares, it could be possible to constrain the location of the gamma-ray production region if it is deep enough inside the BLR.


\section{Conclusions and perspectives}

We find that the gamma-gamma absorption in the BLR is not significant enough to claim discovery for the models of BLR and spectral functions we have investigated. There are hints of absorption in case of 3C 454.3 and TXS 1520+31 with significance of the order of $3\sigma$. An implication of our results could be that the gamma-ray emission zone in FSRQs might be located outside or at the outer edge of the BLR. However, further investigation on binning effects on the SED fits are required. Future work is also expected to improve the modelling of the BLR.


\bigskip 
\begin{acknowledgments}
The \textit{Fermi}-LAT Collaboration acknowledges support for LAT development, operation and data analysis from NASA and DOE (United States), CEA/Irfu and IN2P3/CNRS (France), ASI and INFN (Italy), MEXT, KEK, and JAXA (Japan), and the K.A.~Wallenberg Foundation, the Swedish Research Council and the National Space Board (Sweden). Science analysis support in the operations phase from INAF (Italy) and CNES (France) is also gratefully acknowledged.

Thanks to Pfesesani Van Zyl for providing the PKS~1424-41 dates during radio flares and to Julian Sitarek and the MAGIC Collaboration for providing the MAGIC data points.
\end{acknowledgments}

\bigskip 

\end{document}